 \documentclass[preprint2]{aastex}
 \usepackage{amsmath,alltt}
\newcommand{\bF}{\boldsymbol{F}}
\newcommand{\bG}{\boldsymbol{G}}

\newcommand{\bmu}{\boldsymbol{\mu}}
\newcommand{\bu}{\boldsymbol{u}}
\newcommand{\bw}{\boldsymbol{w}}
\newcommand{\bx}{\boldsymbol{x}}
\newcommand{\DF}{\Delta F}
\newcommand{\Dt}{\Delta t}
\newcommand{\Dx}{\Delta x}
\newcommand{\Domega}{\Delta\omega}
\newcommand{\p}{\partial}
\newcommand{\pF}{\partial F}
\newcommand{\pt}{\partial t}
\newcommand{\pu}{\partial u}
\newcommand{\px}{\partial x}

\slugcomment{}

\shorttitle{CFD}
\shortauthors{Trac \& Pen}

\begin{document}


\title{A PRIMER ON EULERIAN COMPUTATIONAL FLUID DYNAMICS FOR ASTROPHYSICS}


\author{Hy Trac}
\affil{Department of Astronomy and Astrophysics, University of Toronto, Toronto, ON M5S 3H8, Canada}
\email{trac@cita.utoronto.ca}
\and
\author{Ue-Li Pen}
\affil{Canadian Institute for Theoretical Astrophysics, 60 St. George Street, Toronto, ON M5S 3H8, Canada}
\email{pen@cita.utoronto.ca}



\begin{abstract}
We present a pedagogical review of some of the methods employed in Eulerian computational fluid dynamics (CFD).  Fluid mechanics is governed by the Euler equations, which are conservation laws for mass, momentum, and energy.  The standard approach to Eulerian CFD is to divide space into finite volumes or cells and store the cell-averaged values of conserved hydro quantities.  The integral Euler equations are then solved by computing the flux of the mass, momentum, and energy across cell boundaries.  We review both first-order and second-order flux assignment schemes.  All linear schemes are either dispersive or diffusive.  The nonlinear, second-order accurate total variation diminishing (TVD) approach provides high resolution capturing of shocks and prevents unphysical oscillations.  We review the relaxing TVD scheme, a simple and robust method to solve systems of conservation laws like the Euler equations.  

A 3-D relaxing TVD code is applied to the Sedov-Taylor blast wave test.  The propagation of the blast wave is accurately captured and the shock front is sharply resolved.  We apply a 3-D self-gravitating hydro code to simulating the formation of blue straggler stars through stellar mergers and present some numerical results.  A sample 3-D relaxing TVD code is provided in the appendix.

\end{abstract}




\section{Introduction}
Astrophysical structure formation and the dynamics of astrophysical systems involve nonlinear gas dynamical processes which cannot be modeled analytically but require numerical methods.  One would like to address the challenging problem of star formation and how this process produces planetary systems.  Observations of the X-ray emission from hot gas in galaxy clusters, the Sunyaev-Zeldovich effect in the CMB spectrum, and the Lyman alpha forest in the spectra of quasars are only meaningful if we understand the gas dynamical processes involved.  The evolution of complex systems is best modeled using numerical simulations.

A large class of astrophysical problems involve collisional systems where the mean free path is much smaller than all length scales of interest.  Hence, one can appropriately adopt an ideal fluid description of matter where the thermodynamical properties of the fluid obey well known equations of state.  Conservation of mass, momentum, and energy allows one to write down the Euler equations which govern fluid mechanics \citep[See][]{lan87}.  This formalism is an ideal basis for simulating astrophysical fluids.

Hydrodynamical simulations are faced with challenging problems, but advancements in the field have made it an important tool for theoretical astrophysics.  One of the main challenges in simulating complex fluid flows is the capturing of strong shocks, which frequently occur and play an important role in gas dynamics.  The differential Euler equations are ill-defined at shock discontinuities where derivatives are infinite.  Much effort has been devoted to solving this problem and a field of work has resulted from it.  Computational fluid dynamics (CFD) is a powerful approach to simulating fluid flow with emphasis on high resolution capturing of shocks and prevention of numerical instabilities.  Both Eulerian and Lagrangian methods have been developed.
 
Lagrangian methods based on smoothed particle hydrodynamics \citep[SPH;][]{gin77,lucy77} consider a Monte-Carlo approximation to solving the fluid equations, somewhat analogous to $N$-body methods for the Vlasov equation.  SPH schemes follow the trajectories of particles of fixed mass which represent fluid elements.  The Lagrangian forms of the Euler equations are solved to determine smoothed fluid variables like density, velocity, and temperature.  The particle formulation does not naturally capture shocks and artificial viscosity is added to prevent unphysical oscillations.  However, the addition of artificial viscosity broadens shocks over several smoothing lengths and degrades the resolution.  The Lagrangian approach has a large dynamic range in length but not in mass.  It achieves good spatial resolution in high density regions but does poorly in low density regions.  SPH schemes must smooth over a large number of neighbouring particles, making it computationally expensive and challenging to implement in parallel.

The standard approach to Eulerian methods is to discretize the problem and solve the integral Euler equations on a Cartesian grid by computing the flux of mass, momentum, and energy across grid cell boundaries.  In conservative schemes, the flux taken out of one cell is added to the neighbouring cell and this ensures the correct shock propagation.  Flux assignment schemes based on the {\it total variation diminishing condition} \citep{har83} have been designed for high order accuracy and high resolution capturing of shocks, while preventing unphysical oscillations.  The Eulerian approach has a large dynamic range in mass but not in length, opposite to that of Lagrangian schemes.  In general, Eulerian algorithms are computationally faster by several orders of magnitude.  They are also easy to implement and to parallelize.

The purpose of this paper is to present a pedagogical review of some of the methods employed in Eulerian computational fluid dynamics.  In $\S2$ we briefly review the Euler equations and discuss the standard approach to discretizing conservation laws.  We describe traditional central differencing methods such as the Lax-Wendroff scheme in $\S3$ and more modern flux assignment methods like the TVD scheme in $\S4$.  In $\S5$ we review the relaxing TVD method for systems of conservation laws like the Euler equations, which has been successfully implemented for simulating cosmological astrophysical fluids by \citet{pen98}.  In $\S6$ we apply a self-gravitating hydro code to simulating the formation of blue straggler stars through stellar mergers.  A sample 3-D relaxing TVD code is provided in the appendix.

\section{Eulerian Hydrodynamics}

The Euler equations which govern hydrodynamics are a system of conservation laws for mass, momentum, and energy.  In differential conservation form, the continuity equation, momentum equation, and energy equation are given as:
\begin{gather}
\frac{\p\rho}{\pt}+\frac{\p}{\px_j}(\rho v_j)=0\ , \\[8pt]
\frac{\p(\rho v_i)}{\pt}+\frac{\p}{\px_j}(\rho v_iv_j+P\delta_{ij})=0\ , \\[8pt]
\frac{\p e}{\pt}+\frac{\p}{\px_j}[(e+P)v_j]=0\ .
\end{gather}
We have omitted gravitational and other source terms like heating and cooling.  The physical state of the fluid is specified by its density $\rho$, velocity field $\boldsymbol{v}$, and total energy density,
\begin{equation}
e=\frac{1}{2}\rho v^2+\varepsilon\ .
\end{equation}
In practice, the thermal energy $\varepsilon$ is evaluated by subtracting the kinetic energy from the total energy.  For an ideal gas, the pressure $P(\varepsilon)$ is related to the thermal energy by the equation of state,
\begin{equation}
P=(\gamma-1)\varepsilon ,
\end{equation}
where $\gamma$ is the ratio of specific heats.  Another thermodynamic variable which is of importance is the sound speed $c_s$ which is given by
\begin{equation}
c_s^2\equiv\frac{\p P}{\p\rho}=\frac{\gamma P}{\rho}\ .
\end{equation}
The thermodynamical properties of an ideal gas obey well known equations of state, which we do not fully list here.

The differential Euler equations require differentiable solutions and therefore, are ill-defined at jump discontinuities where derivatives are infinite.  In the literature, nondifferentiable solutions are called {\it weak solutions}.  The differential form gives a complete description of the flow in smooth regions, but the integral form is needed to properly describe shock discontinuities.  In integral conservation form, the rate of change in mass, momentum, and energy is equal to the net flux of those conserved quantities through the surface enclosing a control volume.  For simplicity of notation, we will continue to express conservation laws in differential form, as a shorthand for the integral form.

\subsection{Computational Fluid Dynamics}

The standard approach to Eulerian computational fluid dynamics is to discretize time into discrete steps and space into finite volumes or cells, where the conserved quantities are stored.  In the simplest case, the integral Euler equations are solved on a Cartesian cubical lattice by computing the flux of mass, momentum, and energy across cell boundaries in discrete time steps.  Consider the Euler equations in vector differential conservation form,
\begin{equation}
\frac{\p\bu}{\pt}+\frac{\p\bF_i(\bu)}{\px_i}=0\ ,
\end{equation}
where $\bu=(\rho,\rho v_x,\rho v_y,\rho v_z,e)$ contains the conserved physical quantities and $\bF(\bu)$ represents the flux terms.  In practice, the conserved cell-averaged quantities $\bu_n\equiv\bu(\bx_n)$ and fluxes $\bF_n$ are defined at integer grid cell centres $\bx_n$.  The challenge is to use the cell-averaged values to determine the fluxes $\bF_{n+1/2}$ at cell boundaries.  

In the following sections, we describe flux assignment methods designed to solve conservation laws like the Euler equations.  For ease of illustration, we begin by considering a 1-D scalar conservation law,
\begin{equation}
\frac{\pu}{\pt}+\frac{\p F(u)}{\px}=0\ ,
\label{eqn:advect}
\end{equation}
where $F(u)=vu$ and $v$ is a constant advection velocity.  Equation (\ref{eqn:advect}) is referred to as a linear advection equation and has the analytical solution,
\begin{equation}
u(x,t)=u(x-vt,0)\ .
\label{eqn:advectsol}
\end{equation}
The linear advection equation describes the transport of the quantity $u$ at a constant velocity $v$.

In integral flux conservation form, the 1-D scalar conservation law can be written as
\begin{equation}
\frac{\p}{\pt}\int_{x_1}^{x_2}\negthinspace u(x,t)dx+\int_{x_1}^{x_2}\frac{\pF(u)}{\px}dx=0\ ,
\end{equation}
where $x_1\equiv x_{n-1/2}$ and $x_2\equiv x_{n+1/2}$ for our control cells.  Let $F_{n+1/2}^t$ denote the flux of $u$ through cell boundary $x_{n+1/2}$ at time $t$.  Note then that the second integral is simply equal to $F_{n+1/2}^t-F_{n-1/2}^t$.  The rate of change in the cell-integrated quantity $\int\negthinspace udx$ is equal to the net flux of $u$ through the control cell.  For a discrete time step, the discretized solution for the cell-averaged quantity $u_n$ is given by
\begin{equation}
u_n^{t+\Dt}=u_n^t-\left(\frac{F_{n+1/2}^t-F_{n-1/2}^t}{\Dx}\right)\Dt\ .
\label{eqn:conservation}
\end{equation}
The physical quantity $u$ is conserved since the flux taken out of one cell is added to the neighbouring cell which shares the same boundary.  Note that Equation (\ref{eqn:conservation}) has the appearance of being a finite difference scheme for solving the differential form of the 1-D scalar conservation law.  This is why the differential form can be used as a shorthand for the integral form.

\section{Centered Finite-Difference Methods}

Central-space finite-difference methods have ease of implementation but at the cost of lower accuracy and stability.  For illustrative purposes, we start with a simple first-order centered scheme to solve the linear advection equation.  The discretized solution is given by Equation (\ref{eqn:conservation}) where the fluxes at cell boundaries,
\begin{equation}
F_{n+1/2}^t=\frac{F_{n+1}^t+F_n^t}{2}\ ,
\label{eqn:cfdf}
\end{equation}
are obtained by taking an average of cell-centered fluxes $F_n^t=vu_n^t$.  The discretized first-order centered scheme can be equivalently written as 
\begin{equation}
u_n^{t+\Dt}=u_n^t-\left(\frac{F_{n+1}^t-F_{n-1}^t}{2\Dx}\right)\Dt\ .
\label{eqn:cfde}
\end{equation}
In this form, the discretization has the appearance of using a central difference scheme to approximate spatial derivatives.  Hence, centered schemes are often referred to as central difference schemes.  In practice when using centered schemes, the discretization is done on the differential conservation equation rather than the integral equation.

This simple scheme is numerically unstable and we can show this using the {\it von Neumann} linear stability analysis.  Consider writing $u(x,t)$ as a discrete Fourier series:
\begin{equation}
u_n^t=\frac{1}{N}\sum_{k=-N/2}^{N/2}c_k^t\exp\left(\frac{2\pi ikn}{N}\right)\ ,
\label{eqn:dfte}
\end{equation}
where $N$ is the number of cells in our periodic box.  In plane-wave solution form, we can write this as
\begin{equation}
u_n^t=\frac{1}{N}\sum_{k=-N/2}^{N/2}c_k^\circ\exp\left[\frac{2\pi i(kn-\omega t)}{N}\right]\ ,
\end{equation}
where $c_k^\circ$ are the Fourier series coefficients for the initial conditions $u(x,0)$.  Equivalently, the time evolution of the Fourier series coefficients in Equation (\ref{eqn:dfte}) can be cast into a plane-wave solution of the form,
\begin{equation}
c_k^t=\exp\left(\frac{-2\pi i\omega t}{N}\right)c_k^\circ\ ,
\label{eqn:pwck1}
\end{equation}
where the numerical dispersion relation $\omega(k)$ is complex in general.  The imaginary part of $\omega$ represents the growth or decay of the Fourier modes while the real part describes the oscillations.  A numerical scheme is linearly stable if $\text{Im}(\omega)\leq0$.  Otherwise, the Fourier modes will grow exponentially in time and the solution will blow up.

The exact solution to the linear advection equation can be expressed in the form of Equation (\ref{eqn:advectsol}) or by a plane-wave solution where the dispersion relation is given by $\omega_\circ=vk$.  The waves all travel at the same phase velocity $\omega_\circ/k=v$ in the exact case.

The centrally discretized linear advection equation (Equation \ref{eqn:cfde}) is exactly solvable.  After $m$ times steps, the time evolution of the independent Fourier modes is given by
\begin{equation}
c_k^{m\Dt}=\left(1-i\lambda\sin\phi\right)^mc_k^\circ\ ,
\end{equation}
where $\lambda\equiv v\Dt/\Dx$ and $\phi=2\pi k\Dx/N$.  The dispersion relation is given by
\begin{equation}
\omega=\frac{N}{2\pi\Dt}\left[\tan^{-1}(\lambda\sin\phi)+\frac{i}{2}\ln\,(1+\lambda^2\sin^2\phi)\right]\ ,
\end{equation}
For any time step $\Dt>0$, the imaginary part of $\omega$ will be $>0$.  The Fourier modes will grow exponentially in time and the solution will blow up.  Hence, the first-order centered scheme is numerically unstable.

\subsection{Lax-Wendroff Scheme}

The Lax-Wendroff scheme \citep{lax60} is second-order accurate in time and space and the idea behind it is to stabilize the unstable first-order scheme from the previous section.  Consider a Taylor series expansion for $u(x,t+\Dt)$:
\begin{multline}
u(x,t+\Dt)=u(x,t) \\[8pt]
+\frac{\pu}{\pt}\Dt+\frac{\p^2u}{\pt^2}\frac{\Dt^2}{2}+{\cal O}(\Dt^3)\ ,
\end{multline}
and replace the time derivatives with spatial derivatives using the conservation law to obtain
\begin{multline}
u(x,t+\Dt)=u(x,t) \\[8pt]
-\frac{\pF}{\px}\Dt+\frac{\p}{\px}\left(\frac{\pF}{\pu}\frac{\pF}{\pu}\frac{\pu}{\px}\right)\frac{\Dt^2}{2}+{\cal O}(\Dt^3)\ .
\label{eqn:taylor}
\end{multline}
For the linear advection equation, the eigenvalue of the flux Jacobian is $\pF/\pu=v$.  Discretization using central differences gives
\begin{multline}
u_n^{t+\Dt}=u_n^t -\frac{F_{n+1}^t-F_{n-1}^t}{2\Dx}\Dt \\[8pt]
+\left(\frac{F_{n+1}^t-F_n^t}{\Dx}-\frac{F_n^t-F_{n-1}^t}{\Dx}\right)\frac{v\Dt^2}{2\Dx}\ .
\label{eqn:lwe}
\end{multline}
In conservation form, the solution is given by Equation (\ref{eqn:conservation}), where the fluxes at cell boundaries are defined as
\begin{equation}
F_{n+1/2}^t=\frac{1}{2}\left(F_{n+1}^t+F_n^t\right)-\left(F_{n+1}^t-F_n^t\right)\frac{v\Dt}{2\Dx}\ .
\end{equation}
Compare this with the boundary fluxes for the first-order scheme (Equation \ref{eqn:cfdf}).  The Lax-Wendroff scheme obtains second-order fluxes,
\begin{equation}
F^{(2)}=F^{(1)}-\frac{\pF}{\pu}\frac{\pF}{\px}\frac{\Dt}{2}\ ,
\end{equation}
by modifying the first-order fluxes $F^{(1)}$ with a second-order correction.

\begin{figure}[t]
%\plotone{lwdispersion.eps}
\plotone{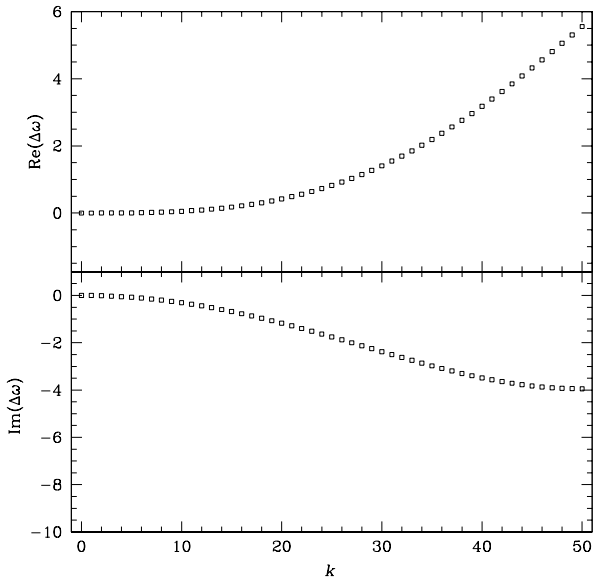}
\caption{The phase error $\text{Re}(\Domega)$ and the amplication factor $\text{Im}(\Domega)$ for the Lax-Wendroff scheme with parameters $N=100$, $v=1$, and $\lambda=0.9$.}
\label{fig:lwdispersion}
\end{figure}

\begin{figure}[t]
%\plotone{laxwendroff.eps}
\plotone{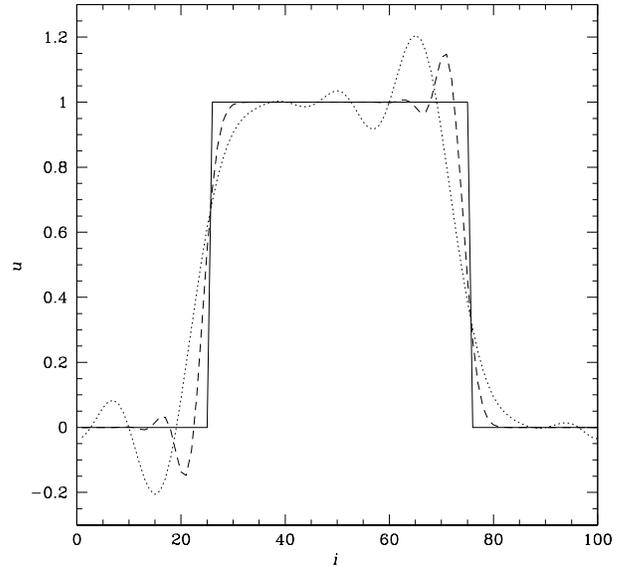}
\caption{Lax-Wendroff scheme used to linearly advect a square wave (solid line) once (dashed line) and ten times (dotted line) through a box of 100 grid cells at speed $v=1$.}
\label{fig:lax}
\end{figure}

The stability of the Lax-Wendroff scheme to solve the linear advection equation can also be determined using the von Neumann analysis.  The discretized Lax-Wendroff equation (Equation \ref{eqn:lwe}) is exactly solvable and after $m$ time steps, the Fourier modes evolve according to
\begin{equation}
c_k^{m\Dt}=\left[1-\lambda^2(1-\cos\phi)-i\lambda\sin\phi\right]^mc_k^\circ\ ,
\label{eqn:lwce}
\end{equation}
where $\lambda\equiv v\Dt/\Dx$ is called the {\it Courant} number and $\phi=2\pi k\Dx/N$.  The dispersion relation is given by 
\begin{multline}
\omega=\frac{N}{2\pi\Dt}\tan^{-1}\left[\frac{\lambda\sin\phi}{1-\lambda^2(1-\cos\phi)}\right] \\[8pt]
+\frac{iN}{4\pi\Dt}\ln\,\left[1-4\lambda^2(1-\lambda^2)\sin^4\left(\frac{\phi}{2}\right)\right]\ .
\label{eqn:lwdr}
\end{multline}
It is important to note three things.  First, the Lax-Wendroff scheme is conditionally stable provided that $\text{Im}(\omega)\leq0$, which is satisfied if
\begin{equation}
\frac{v\Dt}{\Dx}\leq1\ .
\label{eqn:cfl}
\end{equation}
This constraint is a particular example of a general stability constraint known as the {\it Courant-Friedrichs-Lewyor} (CFL) condition.  The Courant number $\lambda$ is also referred to as the CFL number.  Second, for $\lambda=1$ the dispersion relation is exactly identical to that of the exact solution and the numerical advection is exact.  This is a special case, however, and it does not test the ability of the Lax-Wendroff scheme to solve general scalar conservation laws.  Normally, one chooses $\lambda<1$ to satisfy the CFL condition.  Lastly, for $\lambda<1$ the dispersion relation $\omega(k)$ for the Lax-Wendroff solution is different from the exact solution where $\omega_\circ=vk$.  The dispersion relation relative to the exact solution can be parametrized by
\begin{equation}
\Domega\equiv\omega-\omega_\circ\ .
\end{equation}
The second-order truncation of the Taylor series (Equation \ref{eqn:taylor}) results in a phase error $\text{Re}(\Domega)$ which is a function of frequency.  In the Lax-Wendroff solution, the waves are damped and travel at different speeds.  Hence the scheme is both diffusive and dispersive.  

In Figure (\ref{fig:lwdispersion}) we plot the phase error $\text{Re}(\Domega)$ and the amplification term $\text{Im}(\Domega)$ for the Lax-Wendroff scheme with parameters $N=100$, $v=1$, and $\lambda=0.9$.  A negative value of $\text{Re}(\Domega)$ represents a lagging phase error while a positive value indicates a leading phase error.  For the chosen CFL number, the high frequency modes have the largest phase errors but they are highly damped.  Some of the modes having lagging phase errors are not highly damped.  We will subsequently see how this becomes important.

A rigourous test of the 1-D Lax-Wendroff scheme and other flux assignment schemes we will discuss is the linear advection of a square wave.  The challenge is to accurately advect this discontinuous function where the edges mimic Riemann shock fronts.  In Figure (\ref{fig:lax}) we show how the Lax-Wendroff scheme does at advecting the square wave once (dashed line) and ten times (dotted line) through a periodic box of 100 grid cells at speed $v=1$ and $\lambda=0.9$.  Note that this scheme produces numerical oscillations.  Recall that a square wave can be represented by a sum of Fourier or sine waves.  These waves will be damped and disperse when advected using the Lax-Wendroff scheme.  Figure (\ref{fig:lwdispersion}) shows that the modes having lagging phase errors are not damped away.  Hence, the Lax-Wendroff scheme is highly dispersive and the oscillations in Figure (\ref{fig:lax}) are due to dispersion.  We leave it as an exercise for the reader to advect a sine wave using the Lax-Wendroff scheme.  Since there is only one frequency mode in this case, there will be no spurious oscillations due to dispersion, but a phase error will be present.  For a comprehensive discussion on the family of Lax-Wendroff schemes and other centered schemes, see \citet{hir90} and \citet{lan98}.

\section{Upwind Methods}

\begin{figure}[t]
%\plotone{uwdispersion.eps}
\plotone{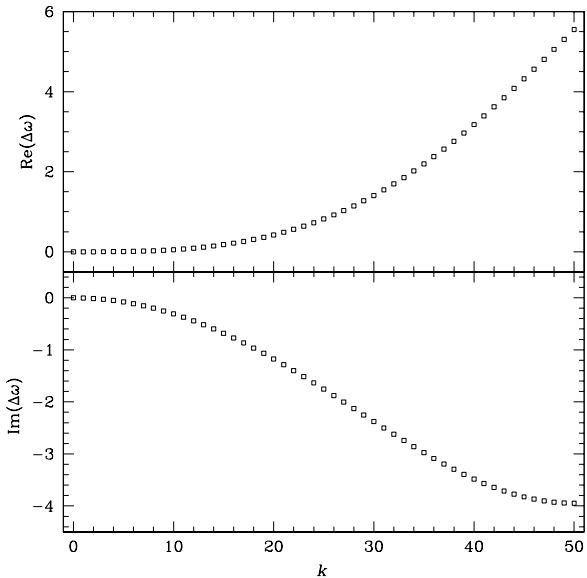}
\caption{The phase error $\text{Re}(\Domega)$ and the amplication term $\text{Im}(\Domega)$ for the Lax-Wendroff scheme (boxes) and the first-order upwind scheme (crosses) with parameters $N=100$, $v=1$, and $\lambda=0.9$.}
\label{fig:uwdispersion}
\end{figure}

\begin{figure}[t]
%\plotone{upwind.eps}
\plotone{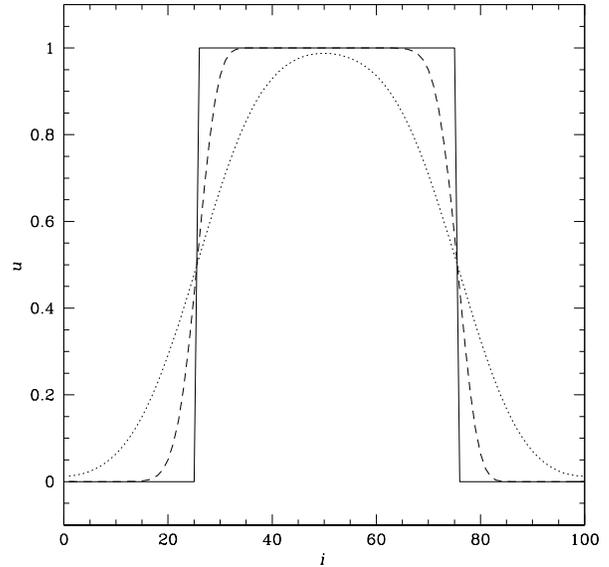}
\caption{First-order upwind scheme used to linearly advect a square wave (solid line) once (dashed line) and ten times (dotted line) through a box of 100 grid cells at speed $v=1$.}
\label{fig:uw}
\end{figure}

Upwind methods take into account the physical nature of the flow when assigning fluxes for the discrete solution.  This class of flux assignment schemes, whose origin dates back to the work of \citet{cou52}, has been shown to be excellent at capturing shocks and also being highly stable.

We start with a simple first-order upwind scheme to solve the linear advection equation.  Consider the case where the advection velocity is positive and flow is to the right.  The flux of the physical quantity $u$ through the cell boundary $x_{n+1/2}$ will originate from cell $n$.  The upwind scheme proposes that, to first-order, the fluxes $F_{n+1/2}^t$ at cell boundaries be taken from the cell-centered fluxes $F_n^t=vu_n^t$, which is in the upwind direction.  If the advection velocity is negative and flow is to the left, the boundary fluxes $F_{n+1/2}^t$ are taken from the cell-centered fluxes $F_{n+1}^t=vu_{n+1}^t$.  The first-order upwind flux assignment scheme can be summarized as follows:
\begin{equation}
F_{n+1/2}^t=
\begin{cases}
F_n^t& \text{if $v>0$ ,} \\[5pt]
F_{n+1}^t& \text{if $v<0$ .}
\end{cases}
\end{equation}
Unlike central difference schemes, upwind schemes are explicitly asymmetric.  

The CFL condition for the first-order upwind scheme can be determined from the von Neumann analysis.  We consider the case of a positive advection velocity.  After $m$ time steps, the Fourier modes evolve according to
\begin{equation}
c_k^{m\Dt}=\left[1-\lambda(1-\cos\phi)-i\lambda\sin\phi\right]^mc_k^\circ\ ,
\label{eqn:uwce}
\end{equation}
where $\lambda\equiv v\Dt/\Dx$ and $\phi=2\pi k\Dx/N$.  The dispersion relation is given by 
\begin{multline}
\omega=\frac{N}{2\pi\Dt}\tan^{-1}\left[\frac{\lambda\sin\phi}{1-\lambda(1-\cos\phi)}\right] \\[8pt]
+\frac{iN}{4\pi\Dt}\ln\,\left[1-4\lambda(1-\lambda)\sin^2\left(\frac{\phi}{2}\right)\right]\ .
\label{eqn:updr}
\end{multline}
The CFL condition for solving the linear advection equation with this scheme is to have $\lambda\leq1$, identical to that for the Lax-Wendroff scheme.  For $\lambda<1$ the dispersion relation $\omega(k)$ for the first-order upwind scheme is different from the exact solution where $\omega_\circ=vk$.  This scheme is both diffusive and dispersive.  Since it is only first-order accurate, the amount of diffusion is large.  In Figure (\ref{fig:uwdispersion}) we compare the dispersion relation of the upwind scheme to that of the Lax-Wendroff scheme.  The Fourier modes in the upwind scheme also have phase errors but they will be damped away.  The low frequency modes which contribute to the oscillations in the Lax-Wendroff solution are more damped in the upwind solution.  Hence, one does not expect to see oscillations resulting from phase errors.

In Figure (\ref{fig:uw}) we show how the first-order upwind scheme does at advecting the Riemann shock wave.  This scheme is well-behaved and produces no spurious oscillations, but since it is only first-order, it is highly diffusive.  The first-order upwind scheme has the property of having monotonicity preservation.  When applied to the linear advection equation, it does not allow the creation of new extrema in the form of spurious oscillations.  The Lax-Wendroff scheme does not have the property of having monotonicity preservation.

The flux assignment schemes that we have discussed so far are all linear schemes.  \citet{god59} showed that all linear schemes are either diffusive or dispersive or a combination of both.  The Lax-Wendroff scheme is highly dispersive while the first-order upwind scheme is highly diffusive.  Godunov's theorem also states that linear monotonicity preserving schemes are only first-order accurate.  In order to obtain higher order accuracy and prevent spurious oscillations, nonlinear schemes are needed to solve conservation laws.

\subsection{Total Variation Diminishing Schemes}

\begin{figure}[t]
%\plotone{minmod.eps}
\plotone{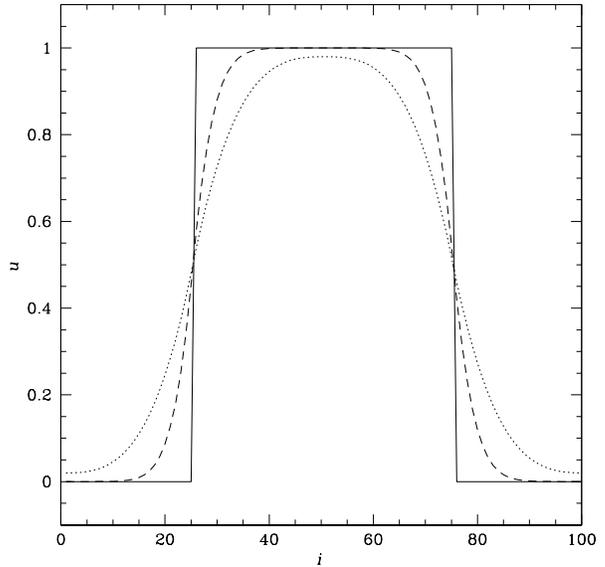}
\caption{The TVD scheme using the minmod flux limiter is applied to the advection of a square wave.}
\label{fig:mm}
\end{figure}

\begin{figure}[t]
%\plotone{superbee.eps}
\plotone{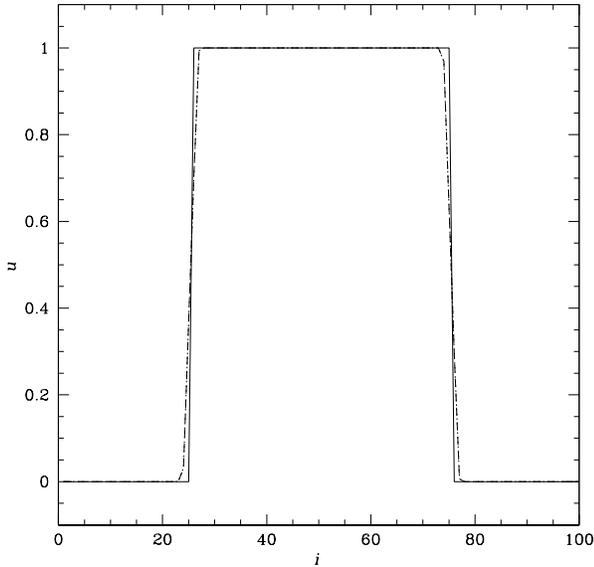}
\caption{The TVD scheme using the superbee flux limiter is applied to the advection of a square wave.}
\label{fig:sb}
\end{figure}

\begin{figure}[t]
%\plotone{vanleer.eps}
\plotone{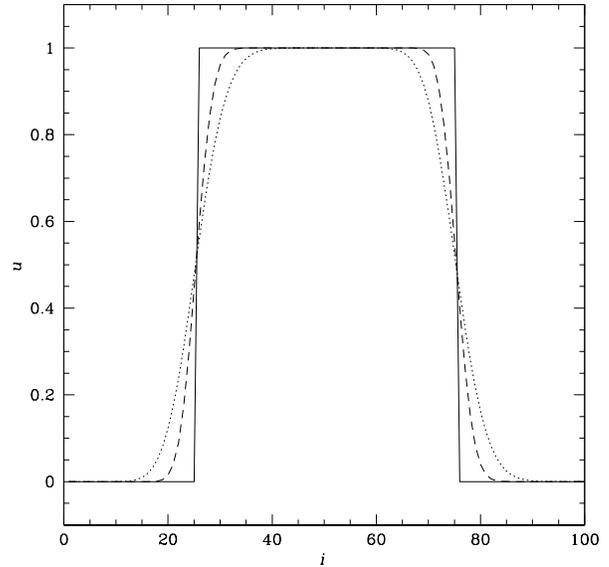}
\caption{The TVD scheme using the Van Leer flux limiter is applied to the advection of a square wave.}
\label{fig:vl}
\end{figure}

\citet{har83} proposed the {\it total variation diminishing} (TVD) condition which guarantees that a scheme have monotonicity preservation.  Applying Godunov's theorem, we know that all linear TVD schemes are only first-order accurate.  In fact, the only linear TVD schemes are the class of first-order upwind schemes.  Therefore, higher order accurate TVD schemes must be nonlinear.

The TVD condition is a nonlinear stability condition.  The total variation of a discrete solution, defined as
\begin{equation}
TV(u^t)=\sum_{i=1}^N|u_{i+1}^t-u_i^t|\ ,
\end{equation}
is a measure of the overall amount oscillations in $u$.  The direct connection between the total variation and the overall amount of oscillations can be seen in the equivalent definition
\begin{equation}
TV(u^t)=2\left(\sum u_{\text{max}}-\sum u_{\text{min}}\right)\ ,
\end{equation}
where each maxima is counted positively twice and each minima counted negatively twice \citep[See][]{lan98}.  The formation of spurious oscillations will contribute new maxima and minima and the total variation will increase.  A flux assignment scheme is said to be TVD if
\begin{equation}
TV(u^{t+\Dt})\leq TV(u^t)\ ,
\end{equation}
which signifies that the overall amount of oscillations is bounded.  In linear flux-assignment schemes, the von Neumann linear stability condition requires that the Fourier modes remain bounded.  In nonlinear schemes, the TVD stability condition requires that the total variation diminishes.

We now describe a nonlinear second-order accurate TVD scheme which builds upon the first-order monotone upwind scheme described in the previous section.  The second-order accurate fluxes $F_{n+1/2}^t$ at cell boundaries are obtained by taking first-order fluxes $F_{n+1/2}^{(1),t}$ from the upwind scheme and modifying it with a second order correction.  First consider the case where the advection velocity is positive.  The first-order upwind flux $F_{n+1/2}^{(1),t}$ comes from the averaged flux $F_n^t$ in cell $n$.  We can define two second-order flux corrections,
\begin{align}
\DF_{n+1/2}^{L,t}&=\frac{F_n^t-F_{n-1}^t}{2}\ ,\\[8pt]
\DF_{n+1/2}^{R,t}&=\frac{F_{n+1}^t-F_n^t}{2}\ ,
\end{align}
using three local cell-centered fluxes.  We use cell $n$ and the cells immediately left and right of it.  If the advection velocity is negative, the first-order upwind flux comes from the averaged flux $F_{n+1}^t$ in cell $n+1$.  In this case, the second-order flux corrections,
\begin{align}
\DF_{n+1/2}^{L,t}&=-\frac{F_{n+1}^t-F_n^t}{2}\ ,\\[8pt]
\DF_{n+1/2}^{R,t}&=-\frac{F_{n+2}^t-F_{n+1}^t}{2}\ .
\end{align}
are based on cell $n+1$ and the cells directly adjacent to it.  Near extrema where the corrections have opposite signs, we impose no second-order correction and the flux assignment scheme reduces to first-order.  A flux limiter $\phi$ is then used to determine the appropriate second-order correction, 
\begin{equation}
\DF_{n+1/2}^t=\phi(\DF_{n+1/2}^{L,t},\DF_{n+1/2}^{R,t})\ ,
\end{equation}
which still maintains the TVD condition.  The second-order correction is added to the first-order fluxes to get second-order fluxes.  The first-order upwind scheme and second-order TVD scheme will be referred to as {\it monotone upwind schemes for conservation laws} (MUSCL).

Time integration is performed using a second-order Runge-Kutta scheme.  We first do a half time step,
\begin{equation}
u_n^{t+\Dt/2}=u_n^t-\left(\frac{F_{n+1/2}^t-F_{n-1/2}^t}{\Dx}\right)\frac{\Dt}{2}\ ,
\end{equation}
 using the first-order upwind scheme to obtain the half-step values $u^{t+\Dt/2}$.  A full time step is then computed, 
\begin{equation}
u_n^{t+\Dt}=u_n^t-\left(\frac{F_{n+1/2}^{t+\Dt/2}-F_{n-1/2}^{t+\Dt/2}}{\Dx}\right)\Dt\ ,
\end{equation}
using the TVD scheme on the half-step fluxes $F^{t+\Dt/2}$.  The reader is encouraged to show that is second-order accurate.

We briefly discuss three TVD limiters.  The minmod flux limiter chooses the smallest absolute value between the left and right corrections:
\begin{equation}
minmod(a,b)=\tfrac{1}{2}[\text{sign}(a)+\text{sign}(b)]\min(|a|,|b|)\ .
\end{equation}
The superbee limiter \citep{roe85} chooses between the larger correction and two times the smaller correction, whichever is smaller in magnitude:
\begin{equation}
superbee(a,b)= 
\begin{cases}
minmod(a,2b)& \text{if }|a|\geq|b|\ ,\\[5pt]
minmod(2a,b)& \text{otherwise .}
\end{cases}
\end{equation}
The Van Leer limiter \citep{vl74} takes the harmonic mean of the left and right corrections:
\begin{equation}
vanleer(a,b)=\frac{2ab}{a+b}\ .
\end{equation}
The minmod limiter is the most moderate of all second-order TVD limiters.  In Figure (\ref{fig:mm}) we see that it does not do much better than first-order upwind for the square wave advection test.  Superbee chooses the maximum correction allowed under the TVD constraint.  It is especially suited for piece-wise linear conditions and is the least diffusive for this particular test, as can be seen in Figure (\ref{fig:sb}).  Note that no additional diffusion can be seen by advecting the square wave more than once through the box.  It can be shown that the minmod and superbee limiters are extreme cases which bound all other second-order TVD limiters.  The Van Leer limiter differs from the previous two in that it is analytic.  This symmetrical approach falls somewhere inbetween the other two limiters in terms of moderation and diffusion, as can be seen in Figure (\ref{fig:vl}).  It can be shown that the CFL condition for the second-order TVD scheme is to have $\lambda<1$.  For a comprehensive discussion on TVD limiters, see \citet{hir90} and \citet{lan98}.

\section{Relaxing TVD}

We now describe a simple and robust method to solve the Euler equations using the monotone upwind scheme for conservation laws (MUSCL) from the previous section.  The relaxing TVD method \citep{jin95} provides high resolution capturing of shocks using computationally inexpensive algorithms which are straightforward to implement and to parallelize.  It has been successfully implemented for simulating cosmological astrophysical fluids by \citet{pen98}.

The MUSCL scheme is straightforward to apply to conservation laws like the advection equation since the velocity alone can be used as a marker of the direction of flow.  However, applying the MUSCL scheme to solve the Euler equations is made difficult by the fact that the momentum and energy fluxes depend on the pressure.  In order to determine the direction upwind of the flow, it becomes necessary to calculate the flux Jacobian eigenvectors using Riemann solvers.  This step requires computationally expensive algorithms.  The relaxing TVD method offers an attractive alternative.

\subsection{1-D Scalar Conservation Law}

We first present a motivation for the relaxing scheme by again considering the 1-D scalar conservation law.  The MUSCL scheme for solving the linear advection equation is explicitly asymmetric in that it depends on the sign of the advection velocity.  We now describe a symmetrical approach which applies to a general advection velocity.  

The flow can be considered as a sum of a right-moving wave $u^R$ and a left-moving wave $u^L$.  For a positive advection velocity, the amplitude of the left-moving wave is zero and for a negative advection velocity, the amplitude of the right-moving wave is zero.  In compact notation, the waves can be defined as:
\begin{align}
u^R&=\left(\frac{1+v/c}{2}\right)u\ ,\\[8pt]
u^L&=\left(\frac{1-v/c}{2}\right)u\ ,
\end{align}
where $c=|v|$.  The two waves are traveling in opposite directions with advection speed $c$ and can be described by the advection equations:
\begin{equation}
\frac{\pu^R}{\pt}+\frac{\p}{\px}(cu^R)=0\ ,
\label{eqn:rmw}
\end{equation}
\begin{equation}
\frac{\pu^L}{\pt}-\frac{\p}{\px}(cu^L)=0\ .
\label{eqn:lmw}
\end{equation}
The MUSCL scheme is straightforward to apply to solve Equations (\ref{eqn:rmw}) and (\ref{eqn:lmw}) since the upwind direction is left for the right-moving wave and right for the left-moving wave.  The 1-D relaxing advection equation then becomes
\begin{equation}
\frac{\pu}{\pt}+\frac{\pF^R}{\px}-\frac{\pF^L}{\px}=0\ .
\label{eqn:rlae}
\end{equation}
where $F^R=cu^R$ and $F^L=cu^L$.  For the discretized solution given by Equation (\ref{eqn:conservation}), the boundary fluxes $F_{n+1/2}^t$ are now a sum of the fluxes $F_{n+1/2}^{R,t}$ and $F_{n+1/2}^{L,t}$ from the right-moving and left-moving waves, respectively.  Note that the relaxing advection equation will correctly reduce to the linear advection equation for any general advection velocity.

Using this symmetrical approach, a general algorithm can be written to solve the linear advection equation for an arbitrary advection velocity.  This scheme is indeed inefficient for solving the linear advection equation since one wave will have zero amplitude.  However, the Euler equations can have both right-moving and left-moving waves with non-zero amplitudes.

\subsection{1-D Systems of Conservation Laws}
\label{sec:1dscl}

We now discuss the 1-D relaxing TVD scheme and later generalize it to higher spatial dimensions.  Consider a 1-D system of conservation laws,
\begin{equation}
\frac{\p\bu}{\pt}+\frac{\p\bF(u)}{\px}=0\ ,
\end{equation}
where for the Euler equations, we have $\bu=(\rho,\rho v,e)$ and $\bF(\bu)$ the corresponding flux terms.  We now replace the vector conservation law with the relaxation system
\begin{equation}
\frac{\p\bu}{\pt}+\frac{\p}{\px}(c\bw)=0\ ,
\label{eqn:relax1}
\end{equation}
\begin{equation}
\frac{\p\bw}{\pt}+\frac{\p}{\px}(c\bu)=0\ ,
\label{eqn:relax2}
\end{equation}
where $c(x,t)$ is a free positive function called the freezing speed.  The relaxation system contains two coupled vector linear advection equations.  In practice, we set $\bw=\bF(\bu)/c$ and use it as an auxiliary vector to calculate fluxes.
Equation (\ref{eqn:relax1}) reduces to our 1-D vector conservation law and Equation (\ref{eqn:relax2}) is a vector conservation law for $\bw$.

In order to solve the relaxed system, we decouple the equations through a change of variables: 
\begin{align}
\bu^R&=\frac{\bu+\bw}{2}\ ,\\[8pt]
\bu^L&=\frac{\bu-\bw}{2}\ ,
\end{align}
which then gives us
\begin{equation}
\frac{\p\bu^R}{\pt}+\frac{\p}{\px}(c\bu^R)=0\ ,
\label{eqn:w1}
\end{equation}
\begin{equation}
\frac{\p\bu^L}{\pt}-\frac{\p}{\px}(c\bu^L)=0\ .
\label{eqn:w2}
\end{equation}
Equations (\ref{eqn:w1}) and (\ref{eqn:w2}) are vector linear advection equations, which can be interpreted as right-moving and left-moving flows with advection speed $c$.  Note the similarity with their scalar counterparts, Equations (\ref{eqn:rmw}) and (\ref{eqn:lmw}).  The 1-D vector relaxing conservation law for $\bu$ becomes
\begin{equation}
\frac{\p\bu}{\pt}+\frac{\p\bF^R}{\px}-\frac{\p\bF^L}{\px}=0\ ,
\end{equation}
where $\bF^R=c\bu^R$ and $\bF^L=c\bu^L$.  The vector relaxing equation can now be solved by applying the MUSCL scheme to Equations (\ref{eqn:w1}) and (\ref{eqn:w2}).  Again, note the similarity between the vector relaxing equation and its scalar counterpart, Equation (\ref{eqn:rlae}).  

The relaxed scheme is TVD under the constraint that the freezing speed $c$ be greater than the characteristic speed given by the largest eigenvalue of the flux Jacobian $\p\bF(\bu)/\p\bu$.  For the Euler equations, this is satisfied for 
\begin{equation}
c=|v|+c_s\ .
\end{equation}
\citet{jin95} considered the freezing speed to be a positive constant in their relaxing scheme while we generalize it to be a positive function.  Time integration is again performed using a second-order Runge-Kutta scheme and the time step is determined by satisfying the CFL condition,
\begin{equation}
\frac{c_\text{max}\Dt}{\Dx}\leq1\ .
\label{eqn:relaxcfl}
\end{equation}
Note that a new freezing speed is computed for each partial step in the Runge-Kutta scheme.  The CFL number $\lambda=c_{max}\Dt/\Dx$ should be chosen such that $c_\text{max}$ will be larger than $\max(c_n^t)$ and $\max(c_n^{t+\Dt/2})$.

We now summarize the steps needed to numerically solve the 1-d Euler equations.  At the beginning of each partial step in the Runge-Kutta time integration scheme, we need to calculate the cell-averaged variables defined at grid cell centres.  First for the half time step, we calculate the fluxes $\bF(u_n^{t})$ and the freezing speed $c_n^t$.  We then set the auxiliary vector $\bw_n^t=\bF(\bu_n^t)/c_n^t$ and construct the right-moving waves $\bu_n^{R,t}$ and left-moving waves $\bu_n^{L,t}$.  The half time step is given by
\begin{equation}
\bu_n^{t+\Dt/2}=\bu_n^t-\left(\frac{\bF_{n+1/2}^t-\bF_{n-1/2}^t}{\Dx}\right)\frac{\Dt}{2}\ ,
\end{equation}
where
\begin{equation}
\bF_{n+1/2}^t=\bF_{n+1/2}^{R,t}-\bF_{n+1/2}^{L,t}\ .
\end{equation}
The first-order upwind scheme is used to compute fluxes at cell boundaries for the right-moving and left-moving waves.  For the full time step, we construct the right-moving waves $\bu_n^{R,t+\Dt/2}$ and left-moving waves $\bu_n^{L,t+\Dt/2}$, using the half-step values of the appropriate variables.  The full time step,
\begin{equation}
\bu_n^{t+\Dt}=\bu_n^t-\left(\frac{\bF_{n+1/2}^{t+\Dt/2}-\bF_{n-1/2}^{t+\Dt/2}}{\Dx}\right)\Dt\ ,
\end{equation}
is computed using the second-order TVD scheme.  This completes the updating of $\bu^t$ to $\bu^{t+\Dt}$.

We have found that a minor modification to the implementation described above gives more accurate results.  Consider writing the flux of the right-moving and left-moving waves as:
\begin{align}
\bF^R=c\bG^R\ ,\\[8pt]
\bF^L=c\bG^L\ ,
\end{align}
where $\bG^R$ is the flux of $\bmu^R=\bu^R/c$ and $\bG^L$ is the flux of $\bmu^L=\bu^L/c$.  The linear advection equations for $\bmu^R$ and $\bmu^L$ are similar to Equations (\ref{eqn:w1}) and (\ref{eqn:w2}), but where we replace $\bu^R$ with $\bmu^R$ and $\bu^L$ with $\bmu^L$.  For each partial step in the Runge-Kutta scheme, the net fluxes at cell boundaries are then taken to be
\begin{equation}
\bF_{n+1/2}=c_{n+1/2}(\bG_{n+1/2}^{R}-\bG_{n+1/2}^{L})\ ,
\end{equation}
where we use $c_{n+1/2}=(c_{n+1}+c_n)/2$.  In practice, this modified implementation has been found to resolve shocks with better accuracy in certain cases.  Note that the two different implementations of the relaxing TVD scheme are identical when a constant freezing speed is used.

\subsection{Multi-Dimensional Systems of Conservation Laws}
\label{sec:mdscl}

The 1-D relaxing TVD scheme can be generalized to higher dimensions using the dimensional splitting technique by \citet{str68}.  In three dimensions, the Euler equations can be dimensionally split into three separate 1-D equations which are solved sequentially.  Let the operator $L_i$ represent the updating of $u^t$ to $u^{t+\Dt}$ by including the flux in the $i$ direction.  We first complete a forward sweep,
\begin{equation}
u^{t+\Dt}=L_zL_yL_xu^t\ ,
\end{equation}
and then perform a reverse sweep
\begin{equation}
u^{t+2\Dt}=L_xL_yL_zu^{t+\Dt}\ ,
\end{equation}
using the same time step $\Dt$ to obtain second-order accuracy.  We will refer to the combination of the forward and reverse sweeps as a double sweep.

A more symmetrical sweeping pattern can be used by permutating the sweeping order when completing the next double time step.  The dimensional splitting or operator splitting technique can be summarized as follows:
\begin{align}
u^{t_2}&=u^{t_1+2\Dt_1}=L_xL_yL_zL_zL_yL_xu^{t_1}\ ,\\[8pt]
u^{t_3}&=u^{t_2+2\Dt_2}=L_zL_xL_yL_yL_xL_zu^{t_2}\ ,\\[8pt]
u^{t_4}&=u^{t_3+2\Dt_3}=L_yL_zL_xL_xL_zL_yu^{t_3}\ ,
\end{align}
where $\Dt_1$, $\Dt_2$, and $\Dt_3$ are newly determined time steps after completing each double sweep. 

The CFL condition for the 3-D relaxing TVD scheme is similarly given by Equation (\ref{eqn:relaxcfl}), but with
\begin{equation}
c_{\max}=\max[(c_x)_{\max},(c_y)_{\max},(c_z)_{\max}]\ .
\end{equation}
where $c_i=|v_i|+c_s$.  Note that since $\max(|v_i|)$ is on average a factor of $\sqrt{3}$ smaller than $\max(|\boldsymbol{v}|)$, a dimensionally split scheme can use a longer time step compared to an un-split scheme.

The dimensional splitting technique also has other advantages.  The decomposition into a 1-D problem allows one to write short 1-D algorithms, which are easy to optimize to be cache efficient.  A 3-D hydro code is straightforward to implement in parallel.  When sweeping in the x direction, for example, one can break up the data into 1-D columns and operate on the independent columns in parallel.  A sample 3-D relaxing TVD code, implemented in parallel using OpenMP directives, is provided in the appendix.

\section{Sedov-Taylor Blast Wave Test for 3-D Hydro}

\begin{figure}[t]
%\plotone{st1.eps}
\plotone{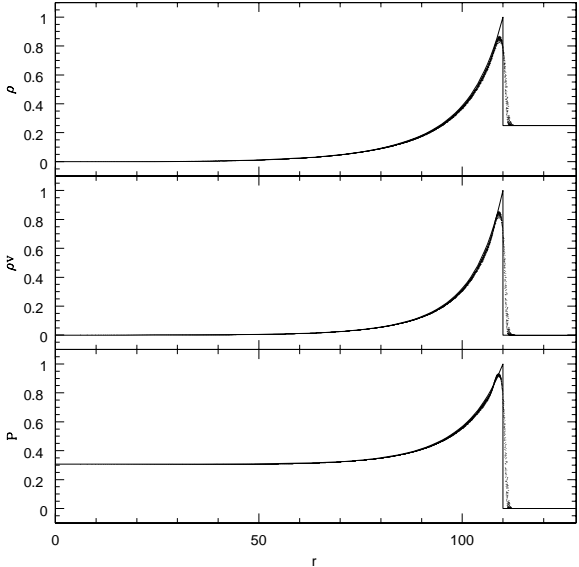}
\caption{Sedov-Taylor blast wave test conducted in a box with $256^3$ cells.  The data points are taken from a random subset of cells and the solid lines are the analytical self-similar solutions.}
\label{fig:st1}
\end{figure}

\begin{figure}[t]
%\plotone{st2.eps}
\plotone{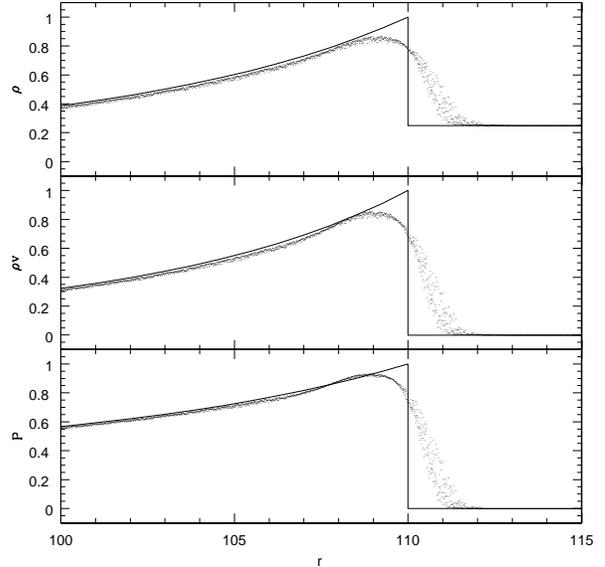}
\caption{A closeup of the Sedov-Taylor blast wave.  The resolution of the shock front is roughly two grid cells and the anisotropic scatter is less than one grid cell.}
\label{fig:st2}
\end{figure}

A rigourous and challenging test for any 3-D Eulerian or Lagrangian hydrodynamic code is the Sedov-Taylor blast wave test.  We set up the simulation box with a homogeneous medium of density $\rho_1$ and negligible pressure and introduce a point-like supply of thermal energy $E_\circ$ in the centre of the box at time $t=0$.  The challenge is to accurately capture the strong spherical shock wave which sweeps along material as it propagates out into the ambient medium.  The Sedov-Taylor test is used to model nuclear-type explosions.  In astrophysics, it is often used as a basic setup to model supernova explosions and the evolution of supernova remnants \citep[See][]{shu92}.

The analytical Sedov solution uses the self-similar nature of the blast wave expansion \citep[See][]{lan87}.  Consider a frame fixed relative to the centre of the explosion.  The spherical shock front propagates outward and the distance from the origin is given by
\begin{equation}
r_{sh}(t)=\xi_\circ\left(\frac{E_\circ t^2}{\rho_1}\right)^{1/5}\ ,
\label{eqn:stre}
\end{equation}
where $\xi_\circ=1.15$ for an ideal gas with $\gamma=5/3$.  The velocity of the shock $v_{sh}=\p r_{sh}/\pt$ is given by
\begin{equation}
v_{sh}(t)=\frac{2}{5}\frac{r_{sh}(t)}{t}\ .
\end{equation}
Since the ambient medium has negligible pressure, the shocks will be very strong.  The density $\rho_2$, velocity $v_2$, and pressure $P_2$ directly behind the shock front are:
\begin{align}
\rho_2&=\left(\frac{\gamma+1}{\gamma-1}\right)\rho_1\ ,\\[8pt]
v_2&=\left(\frac{2}{\gamma+1}\right)v_{sh}\ ,\\[8pt]
\ P_2&=\left(\frac{2}{\gamma+1}\right)\rho_1v_{sh}^2\ .
\end{align}
The immediate post-shock gas density is constant in time, while the shocked gas velocity $v_2$ and pressure $P_2$ decrease as $t^{-3/5}$ and $t^{-6/5}$, respectively.  The full analytical Sedov-Taylor solutions can be found in \citet{lan87}.

The 3-D relaxing TVD code based on the van Leer flux limiter is applied to capturing the Sedov-Taylor blast wave.  We set up a box with $256^3$ cells and constant initial density $\rho_1=1$.  At time $t=0$, we inject a supply of thermal energy $E_\circ=10^5$ into one cell at the centre of the box.  The simulation is stopped at time $t=283$, in which the shock front has propagated out to a distance of $r_{sh}=110$ cells from the centre.  In Figure (\ref{fig:st1}) and (\ref{fig:st2}) we plot the radial distributions of density, momentum, and pressure, normalized to $\rho_2$, $\rho_2v_2$, and $P_2$ respectively.  The data points are taken from a random subset of cells and the solid lines are the analytical Sedov-Taylor solutions.  Despite solving a spherically symmetric problem on an explicitly non-rotationally invariant Cartesian grid, the anisotropic scatter in the results is small.  The distance of the shock front from the centre of the explosion as a function of time is indeed given by Equation (\ref{eqn:stre}), demonstrating that the 3-D relaxing TVD code ensures the correct shock propagation.  The resolution of the shock front is roughly two grid cells.  The numerical shock jump values of $\rho_2$, $v_2$, and $P_2$ are resolution dependent and come close to the theoretical values for our test with $256^3$ cells.  We leave it as an exercise for the reader to test the code using the minmod and superbee flux limiters.

\section{Self-Gravitating Hydro for Astrophysical Applications}

For astrophysical applications, both hydrodynamical and gravitational forces are included.  The gravitational forces arise from the self-gravity of the fluid and can also come from an external field.  The Euler equations with the gravitational source terms included are given as:
\begin{gather}
\frac{\p\rho}{\pt}+\frac{\p}{\px_j}(\rho v_j)=0\ , \\[8pt]
\frac{\p(\rho v_i)}{\pt}+\frac{\p}{\px_j}(\rho v_iv_j+P\delta_{ij})=-\rho\frac{\p\phi}{\px_i}\ , \\[8pt]
\frac{\p e}{\pt}+\frac{\p}{\px_j}[(e+P)v_j]=-\rho v_i\frac{\p\phi}{\px_i}\ .
\end{gather}
where $\phi$ is the gravitational potential.  Poisson's equation,
\begin{equation}
\nabla^2\phi=4\pi G\rho\ ,
\end{equation}
relates the gravitational potential to the density field.  The general solution can be written as
\begin{equation}
\phi(\bx)=\int\rho(\bx')w(\bx-\bx')d^3x'\ ,
\label{eqn:poisson}
\end{equation}
where the kernel is given by
\begin{equation}
w(\bx)=-\frac{G}{|\bx|}\ .
\end{equation}
In the discrete case, the integral in Equation (\ref{eqn:poisson}) becomes a sum and Poisson's equation can be solved using fast Fourier transforms (FFT) to do the convolution.  The forces are then calculated by finite differencing the potential \citep[See][]{he88}.

The addition of gravatitational source terms in the Euler equations is easily handled using the operator splitting technique described in $\S$\ref{sec:mdscl}.  Consider the double sweep:
\begin{equation}
u^{t+2\Dt}=L_xL_yL_zGGL_zL_yL_xu^t\ ,
\end{equation}
where the operator $L_i$ represents the updating of $u$ by including the flux in the $i$ direction and the operator $G$ represents the gravitational acceleration of the fluid.  During the gravitational step, the flux terms in the Euler equations are ignored.  The density distribution does not change and only the fluid momenta and total energy density are updated.

\subsection{Astrophysical Formation of Blue Stragglers Through Stellar Collisions}

The stellar density in the cores of globular and open clusters is high enough for stellar collisions to take place with significant frequency \citep{hil76}.  Current observations and simulations suggest that the merger of two main sequence stars produces a blue straggler \citep{sil97,sbh97}.  The blue stragglers are out-lying main sequence stars which lie beyond the main sequence turnoff in the colour-magnitude diagram (CMD) of a star cluster.  The blue stragglers are more massive, brighter, and bluer than the turnoff stars.  Since more massive stars evolve faster than lower mass stars and are not expected to lie beyond the turnoff, this suggests that blue stragglers must have formed more recently.

In principle the merger of two main sequence stars can produce a young remnant star provided that significant mixing occurs in the process.  The mixing produces a higher hydrogen fraction in the core of the remnant than that of the parent stars which have already burnt most of the hydrogen to helium in their cores.  \citet{bh87} used low resolution SPH simulations with $\sim10^3$ particles to simulate the merging of $n=3/2$ polytropes and found that they fully mixed.  However, medium resolution SPH simulations with $\sim10^4$ particles of $n=3/2$ or $n=3$ polytropes showed only weak mixing \citep{lrs96,sbh97}.  It is worth noting that $n=3/2$ polytropes are more representative of low mass main sequence stars with large convective envelopes while $n=3$ polytropes resemble main sequence stars near the turnoff which have little mass in their convective envelopes.  High resolution SPH simulations involving $\sim10^5-10^6$ particles have now been applied to simulating stellar collisions \citep{sil02}.

The merging stars process is mostly subsonic and strong shocks are not expected.  In the absence of shocks, SPH particles will follow flow lines of constant entropy due to the Lagrangian nature of the method.  As a result, the particles may experience sedimentation.  In addition, the mixing can also depend on the adopted smoothing length and the form of artificial viscosity.  For a SPH fluid, the Reynolds number is of order $(N_p/N_s)^{1/3}$, where $N_p$ is the total number of particles and $N_s$ is the number of particles over which the smoothing is done.  For $N_p\sim10^5$ and $N_s\sim10^2$, the Reynolds number is $\sim10$.  However, a fluid with a low Reynolds number will tend to experience laminar flow.  Hence, SPH may under mix.

It is a worthwhile exercise to model the merging process using Eulerian hydrodynamical simulations.  The differences between Eulerian and Lagrangian approaches may lead to very different results on mixing.  As of present, no such work has been reported in the literature.
 
\subsection{Numerical Method}

\begin{figure}[t]
%\plotone{advectpoly.eps}
\plotone{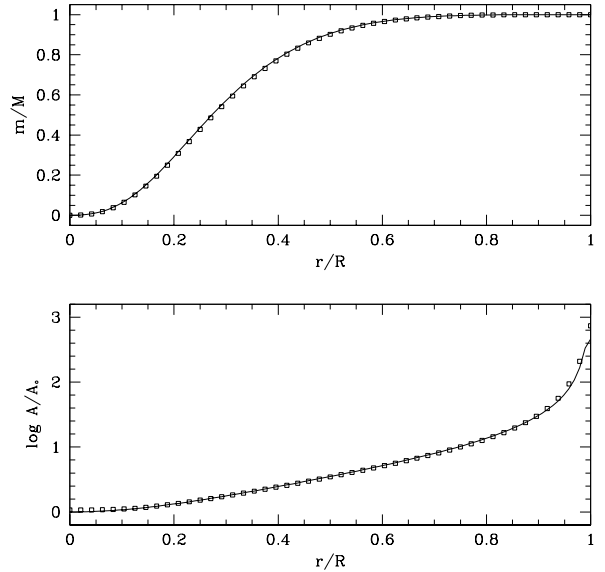}
\caption{The advection of a self-gravitating polytrope in a periodic box with $256^3$ cells.  We compare the mass and entropy profiles of the initial (solid line) and advected polytrope after 1000 timesteps in which the polytrope has moved 256 cells in each direction.}
\label{fig:poly}
\end{figure}

We consider the off-axis collision of two main sequence stars with $M=0.8\, M_\sun$ and $R=0.955\,R_\sun$, which are modeled using $n=3$ polytropes.  A polytrope with polytropic index $n$ has equilibrium density and pressure profiles which are related by
\begin{equation}
P\propto\rho^{1+1/n}\ .
\end{equation}
The density profile is determined by solving the Lane-Emden equation \citep[See][]{cha57}.  We adopt an ideal gas equation of state with adiabatic index $\gamma=5/3$.  Note that for an $n=3$ polytrope, 90\% of the total mass is contained within $r\lesssim0.5R$.  We define the dynamical time to be,
\begin{equation}
\tau_{dyn}\equiv\frac{1}{\sqrt{G\bar{\rho}}}\ ,
%tau_{dyn}\equiv(G\rho)^{-1/2}\ ,
\end{equation}
where $\bar{\rho}$ is the average density.  For the chosen parent stars, the dynamical time is approximately one physical hour.

The collision is simulated in a box with $512\times512\times256$ cells and the orbital plane coincides with the $x-y$ plane.  Initially, each parent star has a radius of 96 grid cells.  The stars are set up on zero-energy parabolic orbits with a pericentre separation equal to $0.25\,R$.  The initial trajectories are calculated assuming point masses.   In an Eulerian simulation, the vacuum cannot have zero density.  We set the minimum density of the cells to be $10^{-8}$ of the central density of the parent stars.  The hydrodynamics is done in a non-periodic box with vacuum boundary conditions.  

\subsection{Numerical Results}

\begin{figure}[t]
%\plotone{densitycontours.eps}
\plotone{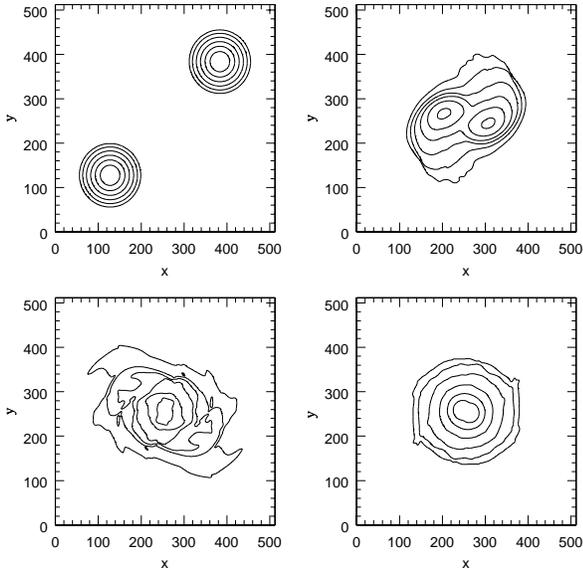}
\caption{Snapshots of the merging process taken at time $t=0$, 2, 4, and $8\,\tau_{dyn}$.  The density contours are spaced logarithmically with 2 per decade and covering 3 decades down from the maximum.}
\label{fig:merge}
\end{figure}

\begin{figure}[t]
%\plotone{profiles.eps}
\plotone{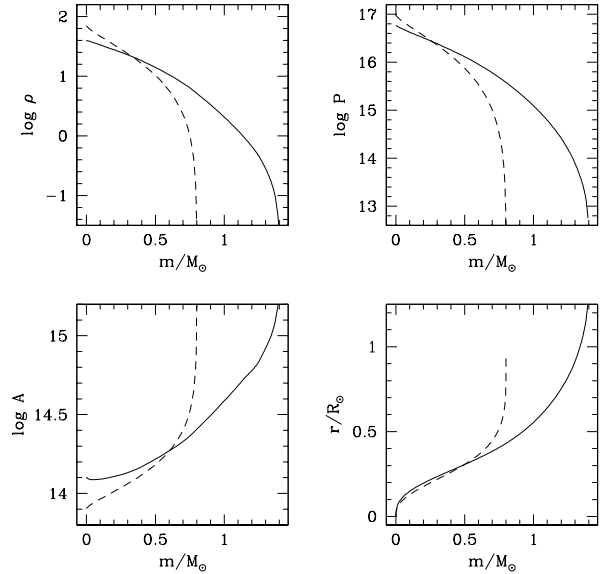}
\caption{Thermodynamic profiles of the merger remnant (solid) and the parent stars (dashed).  Units are cgs.  The radial plot is included for comparison.}
\label{fig:profiles}
\end{figure}

A non-trivial test of a self-gravitating Eulerian hydro code is the advection of an object in hydrostatic equilibrium.  The challenge is to maintain the equilibrium profile over a large number of time steps.  One of the parent stars is placed in a periodic box with $256^3$ cells and given some initial momentum.  We make the test rigorous by having the polytrope move in all three directions.  In Figure \ref{fig:poly} we compare the mass and entropy profiles of the initial and advected polytrope.  The entropic variable $A\equiv P/\rho^\gamma$ is used in place of the specific entropy.  The parameter $A_\circ$ is defined to be the minimum entropy of the parent polytrope.  After 1000 timesteps in which the polytrope has moved 256 cells in each direction, the advected polytrope has still retained its equilibrium profile.  Shock heating can occur in the outer envelope as the polytrope moves through the false vacuum.  However, by setting the density of the false vacuum to be $10^{-8}$ of the central density of the polytrope, we can minimize the spurious shock heating.

In Figure \ref{fig:merge} we show four snapshots of the merging process taken at time $t=0$, 2, 4, and $8\,\tau_{dyn}$.  The 2-D density maps are created by averaging over 4 planes taken about the orbital mid-plane.  The density contours are spaced logarithmically with 2 per decade and covering 3 decades down from the maximum.  The parent stars are initially separated by $3.75\,R$ and placed on zero-energy orbits with a pericentre separation of $0.25R$.  During the collision process, the outer envelopes of the parent stars are shock heated and material gets ejected.  In less than $10\,\tau_{dyn}$, the merger remnant establishes hydrostatic equilibrium.  The merger remnant is a rotating oblate with mass approximately 90\% of the combined mass of the parent stars.  A large fraction of the mass loss is due to the vacuum boundary conditions.  Ejected material do not have the opportunity to fall back onto the merger remnant.  However, the additional mass loss in the envelope does not present a problem since we are interested in the question of mixing in the interior of the star.

In Figure \ref{fig:profiles} we plot the thermodynamic profiles of the merger remnant and the parent stars.  The central density and pressure in the core of the merger remnant is lower than the corresponding values in the parent stars by approximately half.  The entropy floor has risen by a factor of 1.6.  Shock heating is expected to be minimal in the core so a change in entropy suggests that some mixing has taken place.  However, it is difficult to quantify the amount of mixing from examining the thermodynamic profiles alone.

\subsection{Future Work}

To help address the question of mixing, we are implementing a particle-mesh (PM) scheme where test particles can be used to track passively advected quantities such as chemical composition.  Initially, each parent star is assigned a large number of particles with known chemical composition.  The test particles are passively advected along velocity field lines.  For each time step, the velocity of each particle is interpolated from the grid using a "cloud-in-cell" (CIC) scheme \citep{he88} and the equations of motions are solved using second-order Runge-Kutta integration.  The CIC interpolation scheme is also used to determine the local density, pressure, and entropy associated with each particle.  With this setup, we have the benefit of being able to track thermodynamic quantities like in an SPH scheme but avoid the under mixing problem since the fluid equations are solved using the Eulerian scheme.

Future work (Trac, Sills, \& Pen 2003) will have higher resolution simulations.  Collisions will be simulated in a box with $1024\times1024\times512$ cells.  Each parent star will have a radius of 192 grid cells and be assigned $256^3$ test particles.  We will also be doing a detailed comparison between Eulerian and SPH simulations of stellar mergers.

The self-gravitating hydro code used for the simulations is very memory friendly.  For the $1024^2\times512$ grid, 10 GB is required to store the hydro variables, 2 GB for the potential, and less than 1 GB for the test particles.  For every  double timestep, approximately 1000 floating point operations per grid cell is needed to carry out the TVD hydro calculations.  The potential is computed once for every double step and this requires two FFTs.  Since Eulerian codes are very memory friendly, have low floating point counts, are easily parallelized, and scale very well on shared-memory, multiple-processor machines, they can be used to run very high resolution simulations.

\section{Summary}

We have presented several numerical schemes for solving the linear advection equation and given the CFL stability conditions for each scheme.  We have implemented the relaxing TVD scheme to solve the Euler system of conservation laws.  The second-order accurate TVD scheme provides high resolution capturing of shocks, as can be seen in the Riemann shock test and the Sedov-Taylor blast wave test.  The 1-D relaxing TVD scheme can be easily generalized to higher dimensions using the dimensional splitting technique.  A dimensionally split scheme can use longer time steps and is straightforward to implement in parallel.  We have presented a sample astrophysical application.  A 3-D self-gravitating Eulerian hydro code is used to simulate the formation of blue straggler stars through stellar mergers.  We hope to have convinced the reader that Eulerian computational fluid dynamics is a powerful approach to simulating complex fluid flows because it is simple, fast, and accurate.

\acknowledgments

We thank Joachim Stadel and Norm Murray for comments and suggestions on the writing and editing of this paper.  We also thank Alison Sills, Phil Arras, and Chris Matzner for discussions on stellar mergers.

\appendix

\section{3-D Relaxing TVD code}

We provide a sample 3-D relaxing TVD code written in Fortran 90.  The code is implemented using OpenMP directives to run in parallel on shared memory machines.  The code is fast and memory friendly.  The array {\sf u(a,i,j,k)} stores the five conserved hydro quantities ${\sf a}=(\rho,\rho v_x,\rho v_y,\rho v_z,e)$ for each cell {\sf (i,j,k)} in the Cartesian cubical lattice with side length {\sf nc}.  For each sweep, we first call the subroutine {\sf timestep} to determine the appropriate time step {\sf dt} which satisfies the CFL condition.  The updating of {\sf u} by including the flux in the {\sf x} direction is performed by the {\sf sweepx} subroutine.  The data array {\sf u} is divided into 1-D array sections {\sf u1d(a,i)} which are operated on by the {\sf relaxing} subroutine.  The independent columns are distributed amongst multiple processors on a shared memory machine by the OpenMP directives.

The relaxing TVD subroutine in this sample code is written for ease of readability and therefore, is not fully optimized.  At the beginning of each partial step in the Runge-Kutta time integration scheme, the cell-averaged variables defined at grid cell centres are calculated by the {\sf averageflux} subroutine.  The fluxes at cell boundaries for the right-moving and left-moving waves are stored in {\sf fr} and {\sf fl}, respectively.  We have implemented the minmod, superbee, and Van Leer flux limiters and the user of the code can easily switch between them.

We have provided some initial conditions for the Sedov-Taylor blast wave test.  The reader is encouraged to test the code and compare how the various flux limiters do at resolving strong shocks.  This sample code does not implement the modified relaxing TVD scheme described at the end of $\S\ref{sec:1dscl}$, which has been found work very well with the Van Leer flux limiter but unstable with superbee for the 3-D Sedov Taylor test.  We have found that the superbee limiter is often unstable for 3-D fluid simulations.  Please contact the authors regarding any questions on the implementation of the relaxing TVD algorithm.

\begin{alltt}
{\sf
program main
  implicit none
  integer, parameter :: nc=64,hc=nc/2
  real, parameter :: gamma=5./3,cfl=0.9
  
  real, dimension(5,nc,nc,nc) :: u

  integer nsw,stopsim
  real t,tf,dt,E0,rmax

  t=0
  dt=0
  nsw=0
  stopsim=0

  E0=1e5
  rmax=3*hc/4
  tf=sqrt((rmax/1.15)**5/E0)

  call sedovtaylor
  do
     call timestep
     call sweepx
     call sweepy
     call sweepz
     call sweepz
     call sweepy
     call sweepx
     if (stopsim .eq. 1) exit
     call timestep
     call sweepz
     call sweepx
     call sweepy
     call sweepy
     call sweepx
     call sweepz
     if (stopsim .eq. 1) exit
     call timestep
     call sweepy
     call sweepz
     call sweepx
     call sweepx
     call sweepz
     call sweepy
     if (stopsim .eq. 1) exit
  enddo
  call outputresults


contains


  subroutine sedovtaylor
    implicit none
    integer i,j,k

    do k=1,nc
       do j=1,nc
          do i=1,nc
             u(1,i,j,k)=1
             u(2:4,i,j,k)=0
             u(5,i,j,k)=1e-3
          enddo
       enddo
    enddo
    u(5,hc,hc,hc)=E0
    return
  end subroutine sedovtaylor


  subroutine outputresults
    implicit none
    integer i,j,k
    real r,x,y,z

    open(1,file='sedovtaylor.dat',recl=200)
    do k=1,nc
       z=k-hc
       do j=1,nc
          y=j-hc
          do i=1,nc
             x=i-hc
             r=sqrt(x**2+y**2+z**2)
             write(1,*) r,u(:,i,j,k)
          enddo
       enddo
    enddo
    close(1)
    return
  end subroutine outputresults


  subroutine timestep
    implicit none
    integer i,j,k
    real P,cs,cmax
    real v(3)

    cmax=1e-5
    !$omp parallel do default(shared) private(i,j,k,v,cs,P) reduction(max:cmax)
    do k=1,nc
       do j=1,nc
          do i=1,nc
             v=abs(u(2:4,i,j,k)/u(1,i,j,k))
             P=max((gamma-1)*(u(5,i,j,k)-u(1,i,j,k)*sum(v**2)/2),0.)
             cs=sqrt(gamma*P/u(1,i,j,k))
             cmax=max(cmax,maxval(v)+cs)
          enddo
       enddo
    enddo
    !$omp end parallel do
    
    dt=cfl/cmax
    if (t+2*dt .gt. tf) then
       dt=(tf-t)/2
       stopsim=1
    endif
    t=t+2*dt
    nsw=nsw+1
    write(*,"(a7,i3,a8,f7.5,a6,f8.5)") 'nsw = ',nsw,'   dt = ',dt,'   t = ',t
    return
  end subroutine timestep


  subroutine sweepx
    implicit none
    integer j,k
    real u1d(5,nc)

    !$omp parallel do default(shared) private(j,k,u1d)
    do k=1,nc
       do j=1,nc
          u1d=u(:,:,j,k)
          call relaxing(u1d)
          u(:,:,j,k)=u1d
       enddo
    enddo
    !$omp end parallel do
    return
  end subroutine sweepx


  subroutine sweepy
    implicit none
    integer i,k
    real u1d(5,nc)
    
    !$omp parallel do default(shared) private(i,k,u1d)
    do k=1,nc
       do i=1,nc
          u1d((/1,3,2,4,5/),:)=u(:,i,:,k)
          call relaxing(u1d)
          u(:,i,:,k)=u1d((/1,3,2,4,5/),:)
       enddo
    enddo
    !$omp end parallel do
    return
  end subroutine sweepy


  subroutine sweepz
    implicit none
    integer i,j
    real u1d(5,nc)
    
    !$omp parallel do default(shared) private(i,j,u1d)
    do j=1,nc
       do i=1,nc
          u1d((/1,4,3,2,5/),:)=u(:,i,j,:)
          call relaxing(u1d)
          u(:,i,j,:)=u1d((/1,4,3,2,5/),:)
       enddo
    enddo
    !$omp end parallel do
    return
  end subroutine sweepz


  subroutine relaxing(u)
    implicit none
    real, dimension(nc) :: c
    real, dimension(5,nc) :: u,u1,w,fu,fr,fl,dfl,dfr

    !! Do half step using first-order upwind scheme
    call averageflux(u,w,c)
    fr=(u*spread(c,1,5)+w)/2
    fl=cshift(u*spread(c,1,5)-w,1,2)/2
    fu=(fr-fl)
    u1=u-(fu-cshift(fu,-1,2))*dt/2

    !! Do full step using second-order TVD scheme
    call averageflux(u1,w,c)

    !! Right-moving waves
    fr=(u1*spread(c,1,5)+w)/2
    dfl=(fr-cshift(fr,-1,2))/2
    dfr=cshift(dfl,1,2)
    call vanleer(fr,dfl,dfr)
    !call minmod(fr,dfl,dfr)
    !call superbee(fr,dfl,dfr)

    !! Left-moving waves
    fl=cshift(u1*spread(c,1,5)-w,1,2)/2
    dfl=(cshift(fl,-1,2)-fl)/2
    dfr=cshift(dfl,1,2)
    call vanleer(fl,dfl,dfr)
    !call minmod(fl,dfl,dfr)
    !call superbee(fl,dfl,dfr)

    fu=(fr-fl)
    u=u-(fu-cshift(fu,-1,2))*dt
    return
  end subroutine relaxing


  subroutine averageflux(u,w,c)
    implicit none
    integer i
    real P,v
    real u(5,nc),w(5,nc),c(nc)

    !! Calculate cell-centered fluxes and freezing speed
    do i=1,nc
       v=u(2,i)/u(1,i)
       P=max((gamma-1)*(u(5,i)-sum(u(2:4,i)**2)/u(1,i)/2),0.)
       c(i)=abs(v)+max(sqrt(gamma*P/u(1,i)),1e-5)
       w(1,i)=u(1,i)*v
       w(2,i)=(u(2,i)*v+P)
       w(3,i)=u(3,i)*v
       w(4,i)=u(4,i)*v
       w(5,i)=(u(5,i)+P)*v
    enddo
    return
  end subroutine averageflux


  subroutine vanleer(f,a,b)
    implicit none
    real, dimension(5,nc) ::  f,a,b,c

    c=a*b
    where (c .gt. 0)
       f=f+2*c/(a+b)
    endwhere
    return
  end subroutine vanleer


  subroutine minmod(f,a,b)
    implicit none
    real, dimension(nc) :: f,a,b

    f=f+(sign(1.,a)+sign(1.,b))*min(abs(a),abs(b))/2.
    return
  end subroutine minmod


  subroutine superbee(f,a,b)
    implicit none
    real, dimension(5,nc) :: f,a,b

    where (abs(a) .gt. abs(b))
       f=f+(sign(1.,a)+sign(1.,b))*min(abs(a),abs(2*b))/2.
    elsewhere
       f=f+(sign(1.,a)+sign(1.,b))*min(abs(2*a),abs(b))/2.
    endwhere
    return
  end subroutine superbee


end program main
}
\end{alltt}

\end{document}